\documentclass[usegraphicx,usenatbib,useAMS]{mn2e}\usepackage{times}
\usepackage{epstopdf}
 \let\url\relax
\usepackage{graphicx}
\usepackage{amssymb}
\usepackage{caption}
\usepackage{subfig}
\let\url\relax

\def\jcap{{JCAP}}
\def\apj{{ApJ}}
\def\apjs{{ApJs}} 
\def\apjl{{ApJL}}
\def\aap{{A.\&A}}
\def\aj{{AJ}}
\def\pasj{{PASJ}}
\def\prd{{Phys Rev D}}
\def\nat{{Nature}}

\def\MNRAS{{MNRAS}}

\def\mnras{{MNRAS}}
\def\apj{{ApJ}}
\def\apjs{{ApJS}} 
\def\apjl{{ApJL}}
\def\aap{{\em A.\&A}}
\def\aj{{AJ}}
\def\mnras{{MNRAS}}

\def\MNRAS{{MNRAS}}

\def\pasj{{PASJ}}
\def\prd{{Phys Rev D}}
\def\nat{{Nature}}

\newcommand{\be}{\begin{equation}}
\newcommand{\ba}{\begin{eqnarray}}
\newcommand{\ee}{\end{equation}}
\newcommand{\ea}{\end{eqnarray}} 

\def\lesssim{\mathrel{\hbox{\rlap{\hbox{\lower4pt\hbox{$\sim$}}}\hbox{$<$}}}}
\def\gtrsim{\mathrel{\hbox{\rlap{\hbox{\lower4pt\hbox{$\sim$}}}\hbox{$>$}}}}

\def\MNRAS{{\em Mon.\ Not.\ R.\ Astron.\ Soc.\ }}

\def\gtsima{$\; \buildrel > \over \sim \;$}
\def\ltsima{$\; \buildrel < \over \sim \;$}
\def\gsim{\lower.5ex\hbox{\gtsima}}
\def\lsim{\lower.5ex\hbox{\ltsima}}
\def\simgt{\lower.5ex\hbox{\gtsima}}
\def\simlt{\lower.5ex\hbox{\ltsima}}
\def\simpr{\lower.5ex\hbox{\prosima}}
\def\la{\lsim}
\def\ga{\gsim}

\newcommand{\ahf}{\textsc{ahf}}

\def\simless{\mathbin{\lower 3pt\hbox
  {$\rlap{\raise 5pt\hbox{$\char'074$}}\mathchar''7218$}}}  
\def\simgreat{\mathbin{\lower 3pt\hbox
  {$\rlap{\raise 5pt\hbox{$\char'076$}}\mathchar''7218$}}}  

\begin{document}

\title [Extreme objects in Jubilee] {Statistics of extreme objects
in the Juropa Hubble Volume simulation\thanks{http://jubilee-project.org}}
 
\author[W.~A.~Watson, et al.]{William~A.~Watson$^{1}$\thanks{e-mail: 
W.Watson@sussex.ac.uk}, Ilian~T.~Iliev$^{1}$, Jose M. Diego$^{2}$, Stefan Gottl\"ober$^{3}$,
\newauthor Alexander Knebe$^{4}$, Enrique Mart\'inez-Gonz\'alez$^{2}$, Gustavo Yepes$^{4}$
\\
$^1$ Astronomy Centre, Department of Physics \& Astronomy, Pevensey II 
Building, University of Sussex, Falmer, Brighton BN1 9QH, United Kingdom\\  
$^2$ IFCA, Instituto de F\'\i sica de Cantabria (UC-CSIC). Avda. Los Castros s/n. 39005 Santander, Spain\\
$^3$ Leibniz-Institute for Astrophysics, An der Sternwarte 16, 14482, Potsdam, Germany\\
$^4$ Universidad Aut\'onoma de Madrid, Grupo de Astrofisica, 28049, Madrid, Spain\\
}
\date{\today} \pubyear{2013} \volume{000}
\pagerange{1} \twocolumn \maketitle
\label{firstpage}

\begin{abstract}
We present the first results from the JUropa huBbLE volumE (Jubilee) project, based a large N-body, dark matter-only cosmological simulation with a volume of $V=(6~h^{-1}\mathrm{Gpc})^3$, containing 6000$^3$ particles, performed within the concordance $\Lambda$CDM cosmological model. The simulation volume is sufficient to probe extremely large length scales in the universe, whilst at the same time the particle count is high enough so that dark matter haloes down to $1.5\times10^{12}~h^{-1}\mathrm{M}_\odot$ can be resolved. At $z = 0$ we identify over 400 million haloes. The cluster mass function is derived using three different halofinders and compared to fitting functions in the literature. The distribution of clusters of maximal mass across redshifts agrees well with predicted masses of extreme objects, and we explicitly confirm that the Poisson distribution is very good at describing the distribution of rare clusters. The Poisson distribution also matches well the level to which cosmic variance can be expected to affect number counts of high mass clusters. We find that objects like the Bullet cluster exist in the far-tail of the distribution of mergers in terms of relative collisional speed. We  also derive the number counts of voids in the simulation box for $z = 0$, $0.5$ and $1$. 
\end{abstract}

\begin{keywords}
cosmology: large-scale structure of Universe---galaxies: haloes---galaxies: clusters---methods: numerical
\end{keywords}

\section{Introduction}
Surveys mapping a substantial portion of the observable Universe (e.g. BOSS - \citealt{2013AJ....145...10D}, WiggleZ - \citealt{2010MNRAS.401.1429D}, BigBoss - \citealt{2009arXiv0904.0468S}, PanSTARRS - \citealt{2013ApJS..205...20M}, DES - \citealt{2012SPIE.8451E..0DM}, PAU - \citealt{2009ApJ...691..241B}, LSST - \citealt{2012arXiv1211.0310L}, Euclid - \citealt{2012SPIE.8442E..0ZA}, etc.) aim to constrain the cosmological model to unprecedented accuracy. As they will be able to capture very faint objects they will have a shot-noise level low enough to be close to sampling variance-limited. This requires impressive handling of every step of the observational pipeline in order to limit the possibility of systematic errors that may degrade the information contained in them. Aside from these observational efforts, there will also be a similarly high demand placed on our ability to generate theoretical predictions that are equally accurate. This undoubtedly calls for numerical simulations of cosmic structure formation that resolve galactic scales in volumes comparable to the ones covered by these surveys. This is a non-trivial task. A simulation must cover a wide dynamic-range in order to accurately sample large-scale structure (LSS) in the universe. In particular, simulations need to resolve dark matter haloes, which are believed to host the observed galaxies, groups and clusters of galaxies, and accurately model the physics of galaxy formation and other non-linear physics, whilst adequately sampling large-scale matter fluctuations. Only recently have such simulations become feasible and nowadays full-box simulations of considerable fractions of the observable Universe are being conducted utilising close to a trillion particles \citep[for a review of dark matter N-body simulations see ][]{2012PDU.....1...50K}.

Whilst a careful comparison of the statistical clustering properties of objects, in particular galaxies, will put tighter constraints on the parameters of any cosmological model, it is worth noting that the mere existence of individual outliers might pose challenges. Following observations of a series of apparently extreme objects such as high-mass clusters \citep[for example XMMU J2235.3-2557 -- a cluster with mass $\mathrm{M}>4\times 10^{14}~h^{-1}\mathrm{M}_{\odot}$ at redshift $z\sim1.4$ --][]{2005ApJ...623L..85M,2009A&A...508..583R}, or high-velocity collisions (for example the Bullet Cluster -- see \S~\ref{sec:bullet}) some authors have claimed that such objects are highly unlikely to exist in a concordance $\Lambda$CDM cosmology and hence pose a challenge to its validity \citep{2009PhRvD..80l7302J, 2010ApJ...718...60L, 2011MNRAS.415..849C, 2011JCAP...04..017E, 2011PhRvD..83j3502H, 2012ApJ...755L..36H}. However, others have found no tension in their analyses \citep{Harrison2011, 2011PhRvD..83b3015M, 2012A&A...547A..67W} and it has been recently pointed out by \citet{2011JCAP...07..004H} that many studies into high-mass cluster observations have used biased statistical methods, a result that has been corroborated by other studies \citep{2012MNRAS.420.1754W, 2012JCAP...02..009H,2013ApJ...763...93S}. The case of the Bullet cluster is less clear and we discuss it in \S~\ref{sec:bullet}. Debate of this nature highlights the need to clarify our understanding of the statistics of such rare objects, for example by using large cosmological simulations. But this again requires simulations of large enough volumes and sufficient resolution to properly capture the likelihood of the formation of such rare clusters. Theoretical (as opposed to numerical) studies of such objects are challenging too due to their rarity and highly non-linear nature.

In this work we present one of the largest cosmological dark matter only simulations to date, the so-called Jubilee Universe, consisting of  6000$^3$ particles in a cubical volume of side-length 6$~h^{-1}\mathrm{Gpc}$. In this paper we focus on a presentation of the simulation itself and its general properties with respect to the cosmic web, clustering properties, and halo statistics. Subsequent papers will focus on topics including the generation of mock catalogues of LRGs and Quasars, a calculation of the Integrated Sachs-Wolfe effect signal and its cross-correlation to LSS, the weak lensing signal, and the SZ effect \citep[see][for initial results in some of these areas]{2013arXiv1307.1712W}. This paper is laid out as follows. In \S~\ref{sect:meth} we outline our methodology for running the simulation and deriving from it results including halo and void catalogues. In \S~\ref{sect:results} we present our main results and in \S~\ref{sect:disc} briefly discuss their potential implications.

\section{Methods}
\label{sect:meth}
\subsection{Simulations}
\begin{table*}
\caption{N-body simulation parameters. Background cosmology 
is based on the WMAP 5-year results. 
}
\label{summary_N-body_table}
\begin{center}
\begin{tabular}{@{}llllll}
\hline
boxsize & $N_{part}$  & mesh  & smoothing & $m_{particle}$ & $M_{halo,min}$ 
\\[2mm]\hline
3072 $\,h^{-1}$Mpc & $3072^3$ & $6144^3$ & $50~h^{-1}\mathrm{kpc}$ & $7.49\times10^{11}~h^{-1}\mathrm{M}_{\odot}$ & $1.49\times10^{12}~h^{-1}\mathrm{M}_{\odot}$ 
\\[2mm]
6000 $\,h^{-1}$Mpc & $6000^3$ & $12000^3$ & $50~h^{-1}\mathrm{kpc}$ & $7.49\times10^{11}~h^{-1}\mathrm{M}_{\odot}$ & $1.49\times10^{12}~h^{-1}\mathrm{M}_{\odot}$ 
\\[2mm]
\hline
\end{tabular}
\end{center}
\end{table*}

The results presented in this work are based on two large-scale structure N-body simulations, whose parameters are listed in Table~\ref{summary_N-body_table}. Our main simulation has $6000^3$ (216 billion) particles in a volume of  $6\,h^{-1}\mathrm{Gpc}$. The particle mass is $7.49\times10^{11}~h^{-1}\mathrm{M}_{\odot}$, yielding a minimum resolved halo mass (with 20 particles) of $1.49\times10^{12}~h^{-1}\mathrm{M}_{\odot}$, corresponding to galaxies slightly more massive than the Milky Way. Luminous Red Galaxies (LRGs; $\mathrm{M}\sim10^{13}~h^{-1}\mathrm{M}_{\odot}$) are resolved with 100 particles, and galaxy clusters ($\mathrm{M}>10^{14}~h^{-1}\mathrm{M}_{\odot}$) are resolved with $10^3$ particles or more. This main simulation is accompanied by a second, smaller `control' one with $3,072^3$ (29 billion) particles in a volume of $3.072\,h^{-1}\mathrm{Gpc}$ and exactly the same minimum resolution. We used the {\small CUBEP$^3$M} N-body code, a P$^3$M (particle-particle-particle-mesh) code \citep{HarnoisDeraps:2012he}. It calculates the long-range gravity forces on a 2-level mesh and short-range forces exactly, by direct summation over local particles. The code is massively-parallel, using hybrid (combining MPI and OpenMP) parallelization and has been shown to scale well up to tens of thousands of computing cores \citep[see][for complete code description and tests]{HarnoisDeraps:2012he}. Both simulations and most analyses were performed on the Juropa supercomputer at the J\"ulich Supercomputing Centre in Germany (17,664 cores, 53 TB RAM, 207 TFlops peak performance) and required approximately 70,000 and 1.5 million core-hours for the $3\,h^{-1}\mathrm{Gpc}$ and $6\,h^{-1}\mathrm{Gpc}$ boxes respectively. The larger simulation was run on 8,000 computing cores (1,000 MPI processes, each with 8 OpenMP threads), and the smaller one on 2,048 cores.

\subsubsection{Cosmology}

We base our simulation on the 5-year WMAP results \citep{Dunkley:2008ie, Komatsu:2008hk}. The cosmology used was the `Union' combination from \cite{Komatsu:2008hk}, based on results from WMAP, baryonic acoustic oscillations and high-redshift supernovae; i.e.\ $\Omega_{m}=0.27$, $\Omega_{\Lambda}=0.73$, $h=0.70$, $\Omega_{b}=0.044$, $\sigma_8=0.80$, $n_s=0.96$. These parameters are similar to the recent cosmology results of the Planck collaboration \citep{2013arXiv1303.5076P}, where, considering a combination of data from \emph{Planck}, WMAP, and LSS surveys (showing baryon acoustic oscillations) the parameters were calculated to be: $\Omega_{m} = 0.307\pm0.0042$, $\Omega_\Lambda = 0.692\pm0.010$, $h=0.678\pm0.0077$, $\Omega_{b} = 0.048\pm0.00052$, $\sigma_8 = 0.826\pm0.012$ and $n_s = 0.9608\pm0.00024$. The power spectrum and transfer function used for setting initial conditions was generated using CAMB \citep{Lewis:1999bs}. The initial condition generator employed in the run uses first-order Lagrangian perturbation theory (1LPT), i.e.\ the Zel'dovich approximation \citep{1970A&A.....5...84Z}, to place particles in their starting positions. The initial redshift when this step takes place was $z=100$. For a more detailed commentary on the choice of starting redshift for this simulation see \cite{2013MNRAS.433.1230W}.

\subsection{Halofinding}
\label{sect:halofinding}

We use two complementary definitions of haloes in this study. The first is the Spherical Overdensity (SO) definition of \cite{1993MNRAS.262..627L}. In this approach haloes are taken to be spheres that have overdensities that are above a chosen threshold, $\Delta$. The mass enclosed in these spheres is then given by

\be
\label{so_def:eqn}
\mathrm{M}_\Delta = \frac{4\pi \Delta \rho_m}{3} \mathrm{R}^3_\Delta,
\ee

\noindent where $\mathrm{R}_\Delta$ is the radius of the halo and $\rho_m$ is the background matter density in the universe. We choose the overdensity threshold to be $\Delta_{178}$, i.e. an overdensity of 178 times the background matter density. This is a common choice motivated from the top-hat model of non-linear collapse in an Einstein de-Sitter (EdS) universe \citep{1972ApJ...176....1G}. 

The second halo definition we adopt is that of the Friends-of-Friends (FOF) algorithm, first proposed by \cite{1985ApJ...292..371D}. Haloes defined by this algorithm are identified within a simulation volume as agglomerations of particles that lie within a certain parameterised distance from one another. This distance is typically defined as the `linking-length' $\times$ the mean interparticle separation of particles in the simulation. Groups of particles within this  distance of each other are identified as individual dark matter haloes. For our FOF haloes we follow various previous authors \citep{2001MNRAS.321..372J,Reed:2003sq, Reed:2006rw,Crocce:2009mg,Courtin:2010gx,Angulo:2012ep} and use a linking length of 0.2. For further analysis on how the choice of halofinding parameters affect the mass function see \cite{2013MNRAS.433.1230W} and references therein.

We employ three halofinding codes in our analysis: {\small CUBEP$^3$M}'s own on-the-fly SO halofinder (hereafter `CPMSO') \citep{HarnoisDeraps:2012he}, the Amiga Halo Finder \citep[hereafter `AHF',][]{Gill:2004km,Knollmann:2009pb}, and the FOF halofinder from the Gadget-3 N-Body cosmological code \citep[an update to the publicly available Gadget-2 code]{2005MNRAS.364.1105S}.

The CPMSO halofinder utilises a fine mesh from the {\small CUBEP$^3$M} code (with spacing twice as fine as the mean interparticle separation) and an interpolation scheme to identify local peaks in the density field. The code first builds the fine-mesh density using either Cloud-In-Cell (CIC) or Nearest-Grid-Point (NGP) interpolation. It then proceeds to search for and record all local density maxima above a certain threshold (typically set to 100 above the mean density) within the physical volume. It then uses quadratic interpolation on the density field to determine more precisely the location of the maximum within the densest cell. The halo centre determined this way agrees closely with the centre-of-mass of the halo particles. Each of the halo candidates is inspected independently, starting with the highest peak. The grid mass is accumulated in spherical shells of fine grid cells surrounding the maximum until the mean overdensity within the halo drops below $\Delta$. While the mass is accumulated it is removed from the mesh, so that no mass element is double-counted. For further details on the CPMSO method see \cite{HarnoisDeraps:2012he}.

The halofinder \ahf\footnote{\ahf\ is freely available from \texttt{http://www.popia.ft.uam.es/AMIGA}} (\textsc{amiga} Halo Finder) is a spherical overdensity finder that identifies (isolated and sub-)haloes as described in Gill, Knebe \& Gibson (2004) and Knollmann \& Knebe (2009). It employs a recursively refined grid to locate local overdensities in the density field. The identified density peaks are then treated as centres of prospective haloes. The resulting grid hierarchy is further utilised to generate a halo tree containing the information of which halo is a (prospective) host and subhalo, respectively. Halo properties are calculated based on the list of particles asserted to be gravitationally bound to the respective density peak. For a comparison of its performance to other finders in the field we refer the reader to \cite{Knebe:2011rx,2012MNRAS.423.1200O,2013MNRAS.428.2039K}.

The specifics of the FOF halofinder packaged in with the Gadget-3 code currently have not been detailed in any publication but the algorithm itself is outlined in \cite{1985ApJ...292..371D}. The main difference in the algorithm that exists in the Gadget-3 version is that the code is parallelised for distributed-memory machines. Specifically, haloes are found in local subvolumes of the simulation assigned to individual MPI tasks \citep[created using the Gadget-3 domain decomposition which utilises a space-filling Peano-Hilbert curve -- for details see the Gadget-2 paper,][]{2005MNRAS.364.1105S} and then haloes that extend spatially beyond the edges of the subvolumes are linked together in a final MPI communication step. We have altered the Gadget-3 code to read {\small CUBEP$^3$M}'s particle output format and significantly reduced its memory footprint by stripping away extraneous data structures.

Due to limitations in the scaling of the codes with processor numbers and the large memory footprint of the Jubilee simulation it was necessary to split the simulation time-slices into 27 subvolumes and run the halofinding algorithms on each subvolume independently. Each subvolume included a buffer zone which overlapped with the neighbouring ones, for correct handling of haloes straddling two or more sub-regions. We then stitched the subvolumes back together to create the final AHF and FOF halo catalogues, removing any duplicated structures in the overlapping buffers. This approach allows the handling of much larger amounts of data than otherwise possible and provides additional flexibility in terms of computational resources needed for post-processing.

\subsection{Void finding}

The formation of structure in the universe is a hierarchical process: small objects form, grow by accretion and merging and form more and more massive objects up to clusters of galaxies. Between the clusters large filaments can be seen both in observational data as well as in numerical simulations (Figure~\ref{image:fig}). These filaments surround large regions of low density which do not contain objects as massive as the ones found in the filaments or the knots at the end of filaments. These low density regions -- voids -- are the most extended objects in the universe. There are many different ways to define voids and correspondingly there are many different void-finding algorithms \citep[for a review, see][]{2008MNRAS.387..933C}. In the following we are interested in the largest spherical regions of the universe which do not contain any object above a certain threshold in mass. In principle, one could extend this definition of voids also to non-spherical regions, however in this case one can get arbitrary volumes depending on the shape allowed. Since we are interested in the void function we restrict ourselves to spherical voids which are described only by one parameter, their radius. We identify voids in a sample of point-like objects distributed in space. Here these objects are our AHF haloes above a certain mass, but one could also use galaxies above a certain luminosity. Thus our voids are characterised by a threshold mass. If one decreases this mass threshold the number of objects increases and the size of the void decreases. In fact, a given void defined with objects at a higher mass becomes decomposed into many smaller voids defined in the distribution of lower mass objects \citep{2003MNRAS.344..715G}. This reflects the scale-free nature of structure formation. The algorithm searches first for the largest empty sphere then repeats taking into account the previously found voids so that no region is double-counted.

\subsection{Online databases}

It is our intention to make the data from the Jubilee simulation publicly available. This data will consist of three complementary halo catalogues of CPMSO, AHF and FOF haloes, in addition to LRG catalogues derived from the halo data and a catalogue of voids. The CPMSO will be available across a wide number of redshifts ($\sim 30)$ from $z = 0-6$, whereas the AHF and FOF data will be initially available only for $z<1$. Further datasets will include smoothed density fields and maps of weak lensing and ISW signals. An SQL database has been set up so that the data can be queried to suit the requirements of individual users. Further information can be found at the Jubilee project website: http://jubilee-project.org.

\section{Results}
\label{sect:results}
\subsection{Large-scale structure and the Cosmic Web}

\begin{figure*}
 \begin{center}
  \includegraphics[width=7in]{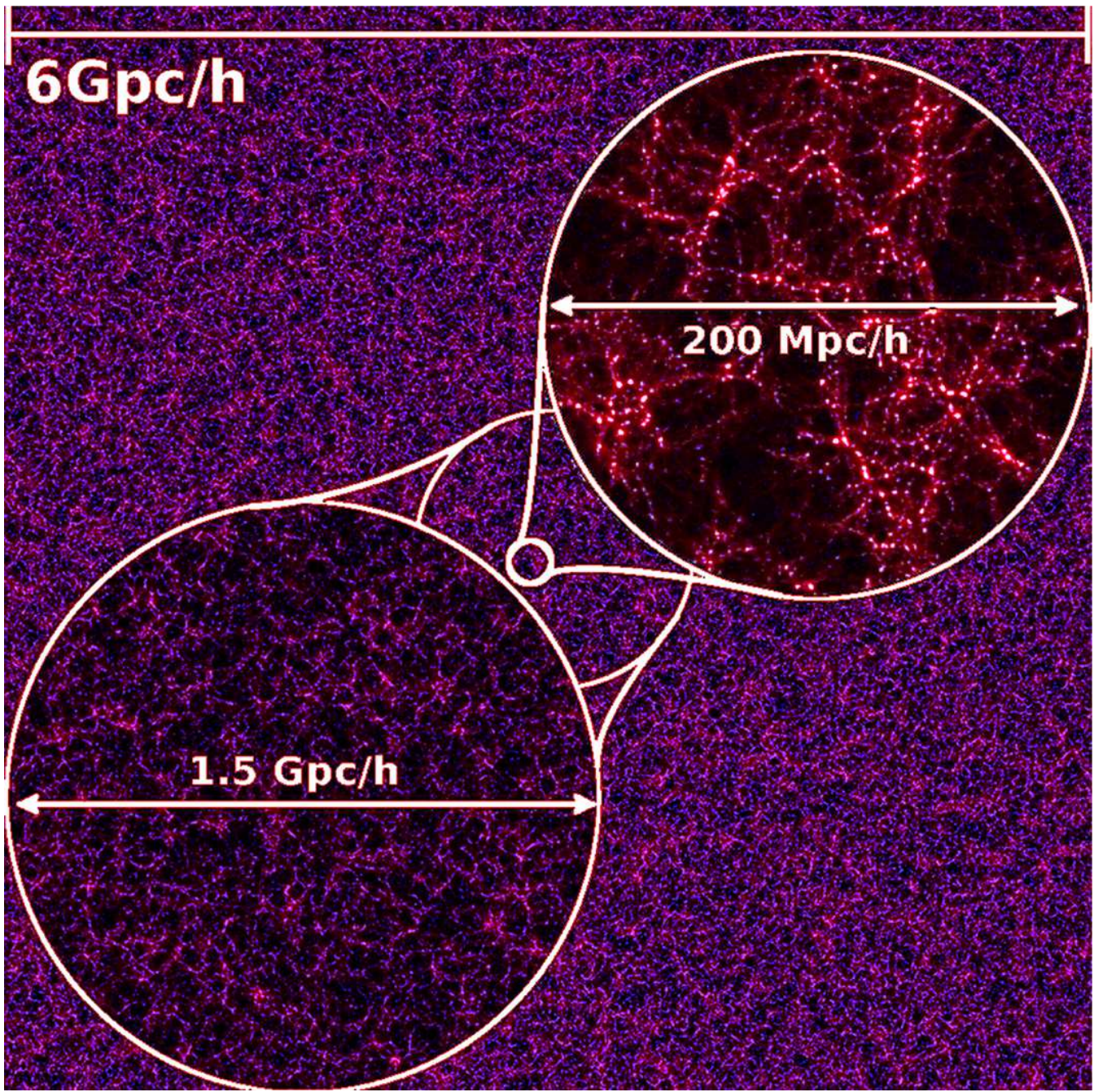}
 \end{center}
 \caption{A slice of the Cosmic Web of structure at $z=0$ based on our $6000^3$-particle simulation. The image is $6~h^{-1}$Gpc per side and $20$~Mpc thick.
  \label{image:fig}}
\end{figure*}

In Figure~\ref{image:fig} we show a slice of the Cosmic Web at $z=0$ extracted from our 6$~h^{-1}\mathrm{Gpc}$ simulation. Perhaps most striking is the homogeneity of the matter distribution at large scales. This is expected from the cosmological principle which states that, on large enough scales, the universe is homogeneous and isotropic. On smaller scales significant features in the density field can be observed including voids, walls, filaments and clusters. For any simulated observations \citep[for example of ISW and weak lensing signals,][]{2013arXiv1307.1712W} we place virtual observers inside the simulation volume at a given location. As can be seen in Figure~\ref{image:fig} the full sky as observed by an observer will show a highly homogeneous distribution of galaxies past a proper distance of a few hundred Mpc (i.e. a redshift of around $z\sim0.1$).

\begin{figure}
 \begin{center}
  \includegraphics[width=3.2in]{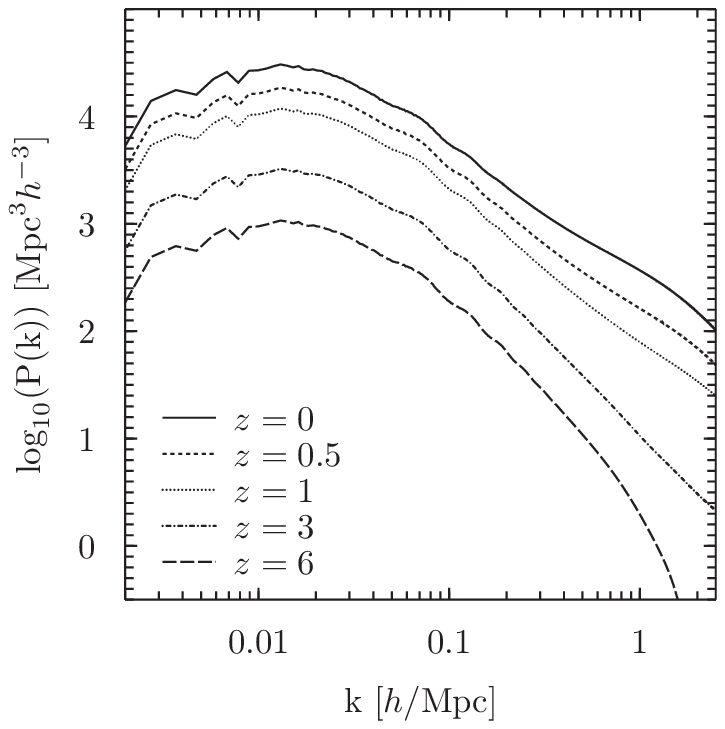}
 \end{center}
 \caption{Evolution of the power spectra of the density field, $P(k)$, as a function of the wavenumber, $k$.
  \label{ps:fig}}
\end{figure}

In Fig.~\ref{ps:fig} we show the evolution of the power spectra of the density field, $P(k)$, from redshift $z=6$ to $z=0$. The particles were interpolated onto a regular grid of $12,000^3$ cells using the cloud-in-cell (CIC) interpolation scheme. From these data we then applied a correction for aliasing and the CIC window function and another for the effect of Poisson noise, all based on the prescription laid out in \cite{2005ApJ...620..559J}.

The baryonic acoustic oscillation (BAO) scale, $k\sim0.1~h\mathrm{Mpc}^{-1}$, is well within the simulation box size and the BAO are visibile in the power spectra. At high redshift, $z\sim6$, the power spectrum is largely linear, except at the smallest scales ($k>1~h\mathrm{Mpc}^{-1}$), where the power grows faster than the linear growth factor predicts. As the hierarchical structure formation proceeds this non-linearity scale propagates to ever larger scales, reaching $k\sim0.1~h\mathrm{Mpc}^{-1}$ at redshift $z=0$, and thereby affecting the BAO scale.

\begin{figure*}
 \begin{center}
  \includegraphics[width=6.5in]{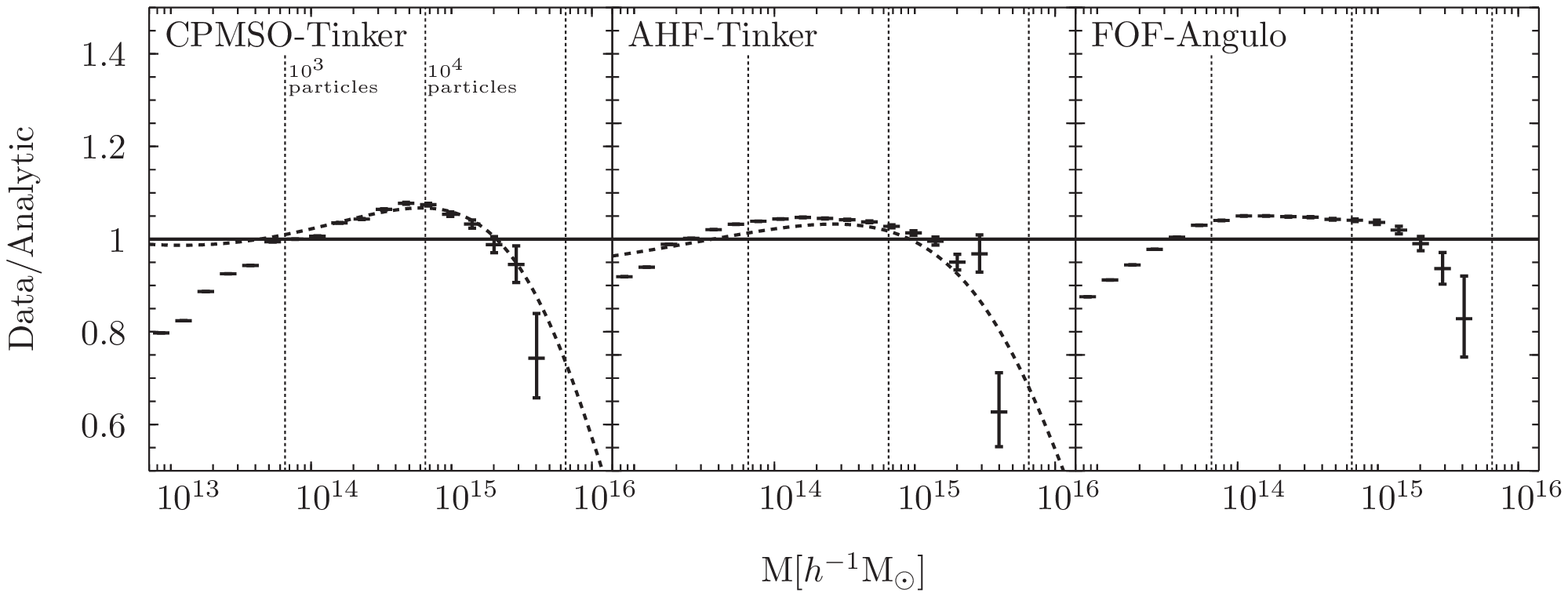}
 \end{center}
 \caption{Jubilee simulation halo mass function based on different halofinders vs. recent analytic fits based on numerical data: (left panel) CPMSO vs. \citet{2008ApJ...688..709T} fit, with the \citet{2013MNRAS.433.1230W} redshift-dependent fit (based on CPMSO haloes) shown as a dashed line; (middle panel) AHF vs. \citet{2008ApJ...688..709T} fit, with the AHF $z = 0$ fit from \citet{2013MNRAS.433.1230W} shown as a dashed line; and (right panel) FOF vs. \citet{Angulo:2012ep}. Errors shown are Poisson.
  \label{mf-fits:fig}}
\end{figure*}

We calculate halo mass functions using the three halofinding algorithms outlined in \S~\ref{sect:halofinding}. Figure~\ref{mf-fits:fig} shows the residuals between our haloes and two fits from the literature at $z = 0$. We compare our CPMSO and AHF haloes to the \cite{2008ApJ...688..709T} mass function, noting that the \cite{2008ApJ...688..709T} fit was calibrated to haloes with an overdensity criteria of $\Delta = 200$ versus the background matter density and our haloes were calculated using a value of $\Delta = 178$. Despite this we see a good correspondence to within $\sim 5\%$ between the \cite{2008ApJ...688..709T} fit and our AHF data for haloes with particle counts greater than $\sim300$. For the very largest haloes there is evidence that the \cite{2008ApJ...688..709T} fit may be overpredicting the mass function, although this is where shot noise begins to severely affect number counts of objects. The CPMSO data follows a similar trend to the AHF data. We overlay on these plots two of the fits from \cite{2013MNRAS.433.1230W} -- mass function results calibrated to data that included the Jubilee haloes presented here. The fit used for the left-hand panel is the redshift-dependent fit based on the CPMSO halofinder, the fit for the central panel a fit based on AHF results for $z=0$ \citep[see][for further details]{2013MNRAS.433.1230W}.

We compare the FOF results to those of the Millennium, Millennium-2 and Millennium-XXL simulations \citep{Angulo:2012ep}, the latter containing $6720^3$ particles in a box with length 3$~h^{-1}$Gpc. The FOF halo data shows agreement to within $\pm5\%$ with the \cite{Angulo:2012ep} fit for haloes with 300 or more particles. The FOF haloes are being compared with a linking length of 0.2, which makes the similarity between the mass functions a good test of the validity of the Jubilee halo distribution as this was the same choice made in \cite{Angulo:2012ep}. For a more detailed study of the mass function across a broad redshift range, including results from the 6$~h^{-1}$Gpc simulation, see \cite{2013MNRAS.433.1230W}.

\subsection{Cosmic variance}
\label{cvar:sec}

\begin{figure*}
 \begin{center}
  \includegraphics[width=5.8in]{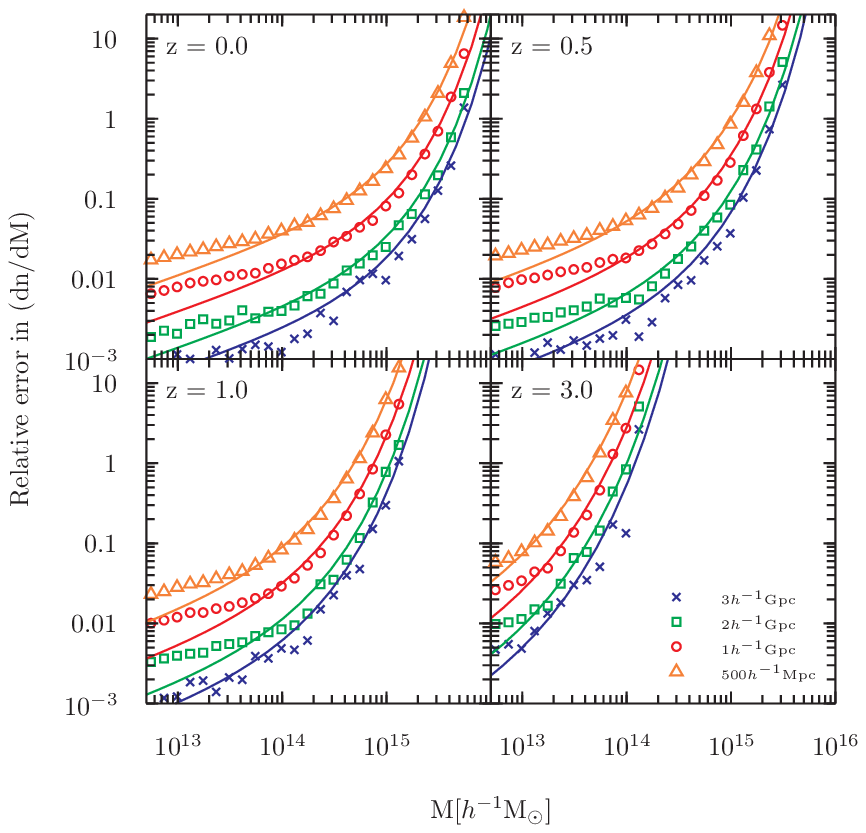}
 \end{center}
 \caption[Relative error in dn/dM for subvolumes of box length 3, 2, 1, and 0.5 $~h^{-1}\mathrm{Gpc}$]{Relative error in dn/dM for subvolumes of box length 3, 2, 1, and 0.5 $~h^{-1}\mathrm{Gpc}$. `Relative error' is defined as the standard deviation from the expected number counts per unit volume in a given mass interval, as defined using the entire $6~h^{-1}\mathrm{Gpc}$ volume. The solid lines are the predictions based on the AHF redshift parameterised mass function from \citet{2013MNRAS.433.1230W} combined with the assumption that the error in number counts around the mean is given by the Poisson distribution.
  \label{cosmic_variance:fig}}
\end{figure*}

Due to the large size of our simulated volume we are able to quantify cosmic variance on scales smaller than our box size in terms of the number counts of objects one expects to find in a given volume. To that end we have compared halo counts in different mass bins in different sized subvolumes. We chose the subvolumes such that they filled the entire full-box with no overlap. The results are shown in Figure~\ref{cosmic_variance:fig}. We show the 1 standard-deviation error in the number counts of haloes by mass bin relative to our entire $(6~h^{-1}\mathrm{Gpc})^3$ volume for sub-box lengths of 3, 2, 1 and 0.5 $~h^{-1}$Gpc. These choices directly compare to the box lengths of some contemporary simulations (Millennium-XXL - \citealt{Angulo:2012ep}, Horizon - \citealt{2009A&A...497..335T}, MultiDark - \citealt{2012MNRAS.423.3018P}, and Millennium - \citealt{2005Natur.435..629S} respectively). We also show a prediction for this error calculated by assuming that the halo number counts follow the CPMSO redshift parameterised mass function from \cite{2013MNRAS.433.1230W}, and by assuming that the observed error in number counts follows a Poisson distribution. This theoretical prediction matches the high-mass data very well. For lower masses the error becomes dominated by sample variance, as discussed in \cite{Smith:2011vm}, and the Poisson prediction presented here begins to break down. We discuss how well the Poisson distribution matches the counts of high mass clusters in \S~\ref{rare:sec} below.

As expected, the error is minimal for lower mass haloes and increases for rarer objects. At $z=0$ the 0.5 $~h^{-1}$Gpc box has an error of under 10\% up until haloes of mass around $4\times10^{14}~h^{-1}\mathrm{M}_{\odot}$, while for box lengths of 1, 2 and 3 $~h^{-1}$Gpc, a 10\% error in number counts per mass interval is realised at around $1\times10^{15},2\times10^{15}$ and $3\times10^{15}~h^{-1}\mathrm{M}_{\odot}$ respectively. The errors are exacerbated at higher redshifts, due to the haloes of a fixed size growing more rare at earlier times.
 
One subtlety should be mentioned: because the sub-volumes considered in this analysis were derived from a larger simulation volume they include the effect of matter fluctuations that exist on scales larger than their box lengths. We stress here that this is not the case for simulations with equivalent volumes to these sub-volumes, as modes of power in the density field that are larger than the box length of a simulation are typically set to zero. This implies that the variation in number counts presented here is slightly different from one that occurs due to a lack of appropriate large-scale power in a simulation volume. This mis-representation of reality (by all simulations, including the Jubilee despite its large volume) leads to an additional set of errors but is, fortuitously, only an issue for very small volume simulations with box lengths of the order of up to a few tens of Mpc \citep[for example see][]{Yoshida:2003wf,2004ApJ...609..474B,Sirko:2005uz,Power:2005ie,Bagla:2006ym,Lukic:2007fc}. Observational volumes, sampling the universe, do not suffer from this effect and the results presented here can be expected to translate reasonably well into counts of high-mass objects in LSS surveys.

\subsection{Statistics of rare objects}
\label{rare:sec}

\begin{figure}
\centering
\includegraphics[width=3.5in]{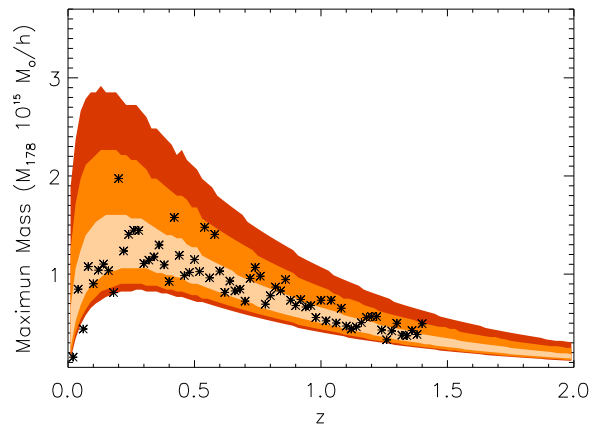}
 \caption{A new version of Figure 1 from \citet{Harrison2012} using the Extreme-Value-Statistics prescription of \citet{Harrison2011} and the CPMSO mass function from \citet{2013MNRAS.433.1230W}. The shaded regions show the 66\%, 95\% and 99\% confidence intervals. The black data points show maximal mass clusters observed by a central observer in the Jubilee simulation.}
  \label{maxmass:fig}
\end{figure}

A current topic in cosmology that relates to the number counts of very high mass objects is that of whether large observed clusters are in conflict with the standard $\Lambda$CDM model. In Figure~\ref{maxmass:fig} we show a theoretical prediction for the expected distribution of maximal mass clusters. This prediction was created using the Extreme Value Statistics (EVS) prescription of \cite{Harrison2011} and the redshift parameterised redshift-dependent mass function, based on CPMSO haloes, from \citet{2013MNRAS.433.1230W}. It is comparable to the plot in Figure 1 of \cite{Harrison2012} which was created using the mass function from \cite{2008ApJ...688..709T}, except for the fact that the mass of the haloes in this version is taken to be set by the $\Delta = 178$ overdensity criterion, rather than the $\Delta = 200$ criterion used in \cite{Harrison2012}. The black data points shown on Figure~\ref{maxmass:fig} correspond to the largest clusters observed in the Jubilee simulation by a central observer, also based on a $\Delta = 178$, as per the configuration of the CPMSO halofinder. The redshift shells for both the EVS contours and the maximal mass clusters are identical. The data all lies within the 3$\sigma$ range showing the expected result that the there is no tension between objects observed in a $\Lambda$CDM cosmological simulation and the theoretical expectation from Extreme Value Statistics. Of interest is whether there are observed clusters in the Universe that have masses that are in tension with the $\Lambda$CDM model. To date observations have shown this to not be the case, as shown in a systematic review by \cite{2012arXiv1210.4369H}.

\begin{figure}
 \begin{center}
  \includegraphics[width=3.5in]{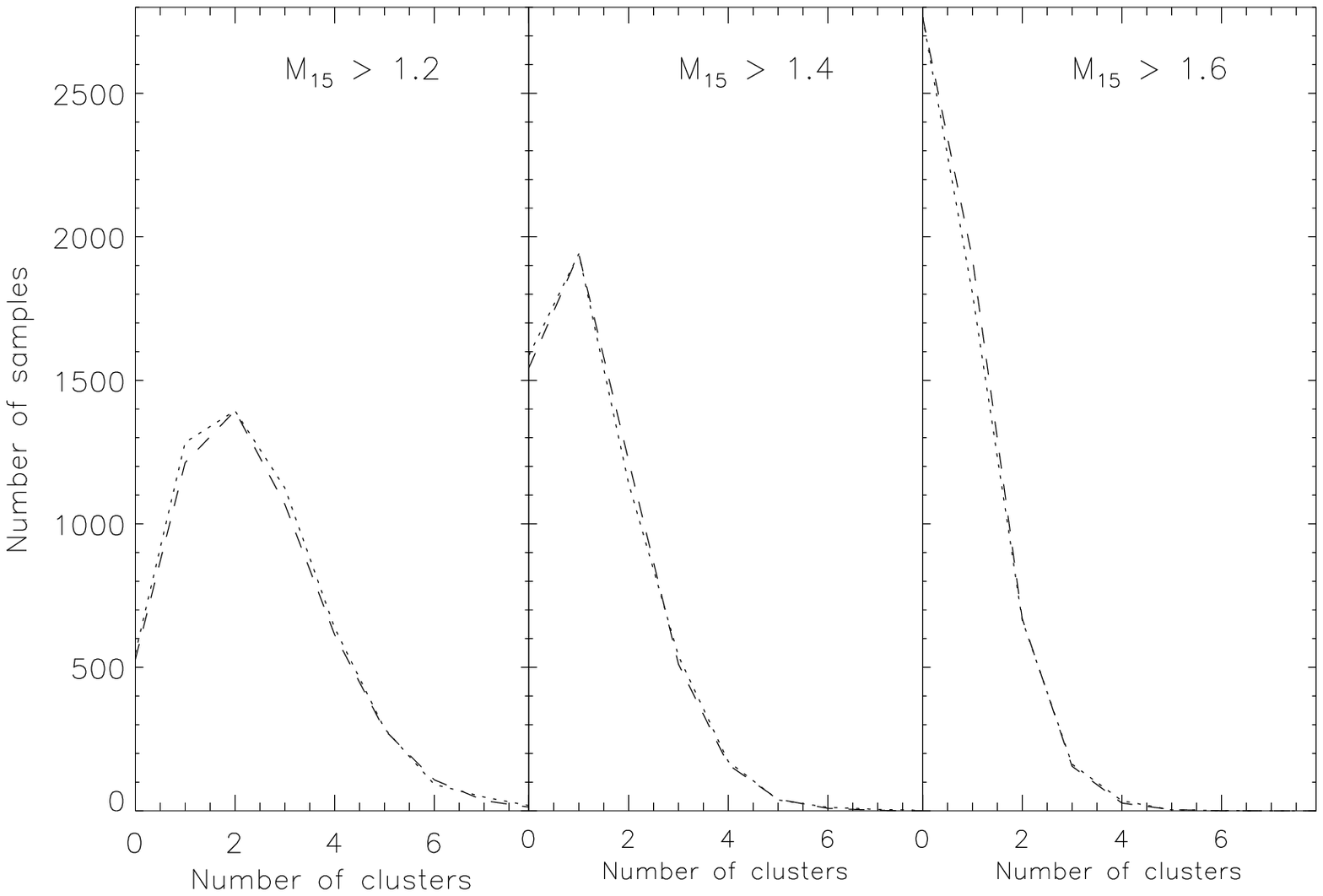}
 \end{center}
 \caption{Histogram of extreme objects for three different thresholds in mass (dotted line) compared with the prediction from the Poisson distribution (dashed line) for the corresponding mean value of the objects above the corresponding threshold. The mass thresholds are: $1.2\times10^{15}~h^{-1}\mathrm{M}_{\odot}$, $1.4\times10^{15}~h^{-1}\mathrm{M}_{\odot}$ and $1.6\times10^{15}~h^{-1}\mathrm{M}_{\odot}$ left-to-right panels respectively. The statistics are calculated for $z=0.05$.
  \label{poisson:fig}}
\end{figure}

The question of how well the Poisson distribution fits our rare cluster number counts is addressed in Figure~\ref{poisson:fig}. The simulation volume at $z=0.05$ was split up into 5438 independent subvolumes. For each subvolume we calculated the number of objects above a given threshold mass ($1.2\times10^{15}~h^{-1}\mathrm{M}_{\odot}$, $1.4\times10^{15}~h^{-1}\mathrm{M}_{\odot}$ and $1.6\times10^{15}~h^{-1}\mathrm{M}_{\odot}$ for the panels in Figure~\ref{poisson:fig}, left-to-right respectively) found in each subvolume. The mass thresholds were chosen so that only a very small number (around 0--2) of objects were found in each subvolume, which represents the regime where we expect Poisson statistics to be dominant. We then compared the histogram of the measured distribution of the objects in the simulation to that predicted by a Poisson distribution with a mean set by the average across all the subvolumes. The correspondence between the two is very close. This is an interesting result as it validates the common choice of Poisson statistics for describing the expected distribution of these objects, and is the first time it has been validated using a simulation of this scale \citep[for a detailed investigation of the applicability of the Poisson distribution in cluster counts across different masses see][who used simulations of box length $1.5~h^{-1}\mathrm{Gpc}$ for their study]{Smith:2011vm}.

\subsection{High $\Delta v$ Mergers and the Bullet Cluster}
\label{sec:bullet}

There has been recent debate regarding whether the Bullet Cluster (1E0657-56, which resides at a redshift of $z=0.296$) poses a challenge to the $\Lambda$CDM model. 1E0657-56 consists of a large cluster of mass $\mathrm{M}_{200}\sim1.5\times10^{15}~h^{-1}\mathrm{M}_{\odot}$ and a sub-cluster -- the `bullet' -- of mass $\mathrm{M}_{200}\sim1.5\times10^{14}~h^{-1}\mathrm{M}_{\odot}$ that has traversed through the larger cluster, creating a substantial bow shock along the way \citep{2002ApJ...567L..27M,2002A&A...386..816B,2004ApJ...604..596C,2006ApJ...648L.109C,2006ApJ...652..937B}. Tension with $\Lambda$CDM arises from the calculated value for the speed of the shock of $v_s = 4740^{+710}_{-550}~\mathrm{kms}^{-1}$ \citep{2002ApJ...567L..27M,Markevitch2004,Springel2007}, which was originally calculated to be too high for a $\Lambda$CDM universe \citep{2007PhRvL..98q1302F} -- whereas it might be better accommodated in alternative cosmologies \citep[e.g.][]{2009ApJ...695L.145L}. Other studies have concluded that the velocity is not in tension with $\Lambda$CDM \citep{2006MNRAS.370L..38H}. An important clarification of this issue was presented by groups working on simulations of Bullet-like systems \citep{2005ApJ...629..791T,2006PASJ...58..925T,2007ApJ...661L.131M,Springel2007,2008MNRAS.389..967M} where, in general, it was found that the shock speed was substantially higher than the speed of the mass centroid of the infalling subcluster. For example \cite{Springel2007} found that a Bullet-like system in their simulations had a shock speed of $\sim4500~\mathrm{kms}^{-1}$ whereas the sub-cluster had a speed of only $\sim2600~\mathrm{kms}^{-1}$. \cite{2007ApJ...661L.131M} found that in an illustrative simulation the sub-cluster CDM halo had a speed that was 16\% lower than that of the shock.

Even given this moderation of the extreme sub-cluster speed in 1E0657-56 there have still been claims in the literature that the $\Lambda$CDM model may be incapable of creating such a system \citep{2010ApJ...718...60L,2012MNRAS.419.3560T}. This is not wholly unexpected as a) \cite{2008MNRAS.389..967M} have shown that the properties of the bow shock are not well described by simulations and b) even with a moderation in sub-cluster speed along the lines of \cite{Springel2007} or \cite{2007ApJ...661L.131M}, the speed may still be too high for the $\Lambda$CDM model to accommodate. These studies have relied on numerical simulations to observe the distribution of relative velocities in colliding clusters. From these distributions 1E0657-56 can be assessed and deemed to be either rare for a $\Lambda$CDM universe or so rare that it puts the whole model in doubt.

Alternative approaches have also been taken in addressing this question. \cite{2010ApJ...725..598F} looked in 2-D-projected position-space for Bullet-like systems in the MareNostrum Universe, a large hydrodynamical cosmological simulation. The characteristic distribution of gas and dark matter in 1E0657-56, as projected on the sky -- with a large displacement between the cluster's gas and dark matter -- was found to be expected in 1\% - 2\% of clusters with masses larger than $10^{14}~h^{-1}\mathrm{M}_{\odot}$. \cite{2008MNRAS.384..343N} performed a `back in time' analysis to place bounds on the relative overdensity the system resides in the universe, concluding that for a relative speed of $\sim4500~\mathrm{kms}^{-1}$ the system would need to have a mass of $2.8\times10^{15}~h^{-1}\mathrm{M}_{\odot}$ and exist in a local overdensity of 10 times the background density of the universe.

\begin{figure*}
 \begin{center}
  \includegraphics[width=5in]{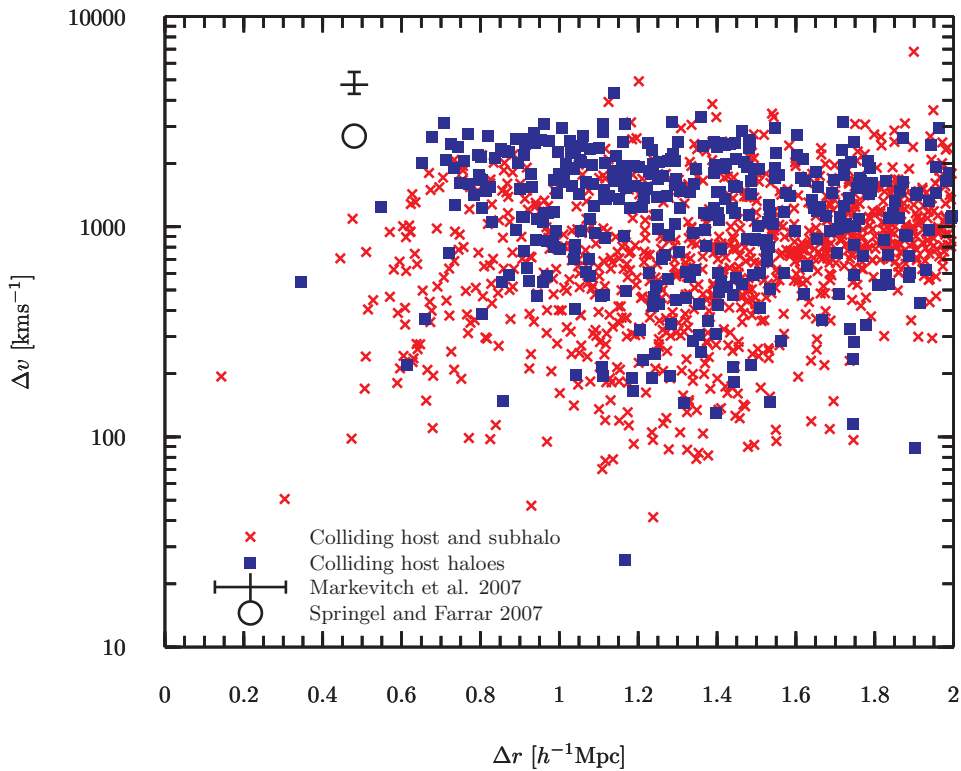}
 \end{center}
 \caption{Relative pairwise velocities for haloes. The data points correspond to the observed bullet cluster speed from \citet{Markevitch2004} (cross) and the corrected speed estimated by \citet{Springel2007} (circle). The simulated speeds were obtained from the full box using the output redshift slice ($z = 0.32$) that best matched the redshift of the bullet cluster $(z = 0.296)$. The halofinder used was AHF, which can resolve both subhaloes and haloes. We have shown in blue merging objects that are both host haloes in their own rights, and in red merging objects where a subhalo is moving with a high relative velocity to its host halo. The data in the simulation was reduced by considering haloes with a mass of over $1\times10^{13}~h^{-1}\mathrm{M}_{\odot}$ and data points were plotted with the restriction that one of the haloes in the pair had a mass of at least $7\times10^{14}~h^{-1}\mathrm{M}_{\odot}$. Finally, we have added in a random scatter to mimic the effect of Eddington bias, as described in the text.}
  \label{bullet:fig}
\end{figure*}

Here we use the huge number counts of clusters in the Jubilee simulation to add to the debate. We consider AHF (sub-)haloes with mass greater than $1\times10^{13}~h^{-1}\mathrm{M}_{\odot}$ that are colliding with (host) haloes of mass greater or equal to $7\times10^{14}~h^{-1}\mathrm{M}_{\odot}$ at $z = 0.32$. Our results are shown in Figure~\ref{bullet:fig}, along with the original Bullet speed presented in \cite{Markevitch2004}, and the moderated result from \cite{Springel2007}, which represents the lowest value from the literature to-date. We show in blue haloes that are colliding pairs and in red haloes that are a colliding subhalo and halo pair. We have added a normally distributed random scatter to our velocities with a width given by the error in the observed value for $v_s$. This is to mimic the effect of Eddington bias in our simulated data. This observational bias arises when observing extreme measurements in a distribution of measurements all with some associated scatter. As there are many more data points that exist with less extreme velocities than the one in question it is likely that an extreme data point is an upscattered less extreme one. As we know very precisely the pairwise velocities of halo pairs in our simulation adding random scatter to this distribution serves to create the effect in our measurement.

As can be seen from the distribution the Bullet cluster is an extreme object, but only when the radial separation of the halo pairs is considered. We find many candidate mergers in our volume with a collision speed that equals or exceeds the more conservative speed estimate for the cluster, and a few objects that are not far from the higher velocity estimate of \cite{Markevitch2004}. However, we find no objects that, at a closer separation, give rise to a large enough merging velocity. This is likely to be due to the effect of only considering one simulation output in our analysis. At any given output redshift only a handful of haloes will be undergoing a major merger event of the sort we are interested in and this is reflected by the paucity of data points that lie at a separation of less than 0.6~$h^{-1}$Mpc. It is likely, therefore, that over the course of the Jubilee simulation run, high velocity mergers of the type observed in the Bullet Cluster do occur. 

This result is in line with previous attempts to use large cosmological simulations to address this issue where the bullet was not found to be extreme \citep{2006MNRAS.370L..38H,2012MNRAS.419.3560T}. Interestingly, \cite{2012MNRAS.419.3560T} extrapolated their results from smaller simulation volumes and concluded that a volume of $(4.5~h^{-1}\mathrm{Gpc})^3$ would be required in order to observe a Bullet-like cluster.

The conclusion that we put forward based on this result is that there is at present no tension between our data and the standard cosmological model. This conclusion is tentative, however, and there would appear to be a need for careful further research into this question based on a number of points. Firstly, the results are very sensitive to the mass cuts imposed on the candidate search. It would be very difficult to find a precise analogue to the Bullet Cluster in terms of the masses, velocities, and spatial separation of the haloes. Here we have taken a cut-off in mass that allows us to search for Bullet-like systems rather than a precise Bullet Cluster. Secondly, we have placed no restrictions on the directions of the relative velocities of the halo pairs. The bow shock observed in the Bullet Cluster has arisen from the Bullet subhalo having passed through the parent halo (it is this occurrence, that fortuitously lies almost in the plane of the sky as we observe it, that has allowed us to identify the relative pairwise velocities of the halo and subhalo in the system). In our analysis we plot \textit{all} the pairwise velocities of the haloes, making no distinction between haloes that are infalling and haloes that have already undergone a collision and not considering how the orientations of the collisions might appear to a specific observer. This is a fair way to assess the data as the actual collision in a Bullet Cluster-like system is expected to take a few hundred Megayears so is a relatively short event. Canvasing all our haloes in this manner assesses whether there is likely to be or whether there has been a Bullet Cluster-like collision in the simulation around $z\sim0.3$. Further studies should include how random observers would observe these events. Lastly, halofinding algorithms are notoriously sub-optimal when trying to find and separate haloes that are merging \citep[this is discussed in detail in][we draw the reader's attention to Figure 10 from that paper in particular]{Knebe:2011rx}. However, as the separation in question of the two haloes is relatively large, this is unlikely to be affecting our results.

\subsection{The Jubilee void function}
\label{void:sec}

The distribution of voids for a given threshold is characterised by the void function, the number of spheres with radii larger than $R_{\mathrm{void}}$ per volume. We have studied the void distribution at redshifts $z = 0$, 0.5, and 1. At $z = 0$ we have identified the voids in the distribution of haloes more massive than $5\times10^{14}~h^{-1}\mathrm{M}_{\odot}$, $2\times10^{14}~h^{-1}\mathrm{M}_{\odot}$, $1\times10^{14}~h^{-1}\mathrm{M}_{\odot}$, and $1\times10^{13}~h^{-1}\mathrm{M}_{\odot}$. At redshift $z = 0$ we identified 244,989, 1,753,982, 5,596,627, 91,615,821 haloes more massive than these thresholds, respectively. Thus the mean distance between them (i.e. the box length divided by the cube root of the number counts) is about 96$~h^{-1}$Mpc, 50$~h^{-1}$Mpc, 34$~h^{-1}$Mpc, and 13$~h^{-1}$Mpc. Nevertheless, we found huge volumes which do not contain any of these objects.

\begin{figure}
\begin{center}
  \includegraphics[width=3.3in]{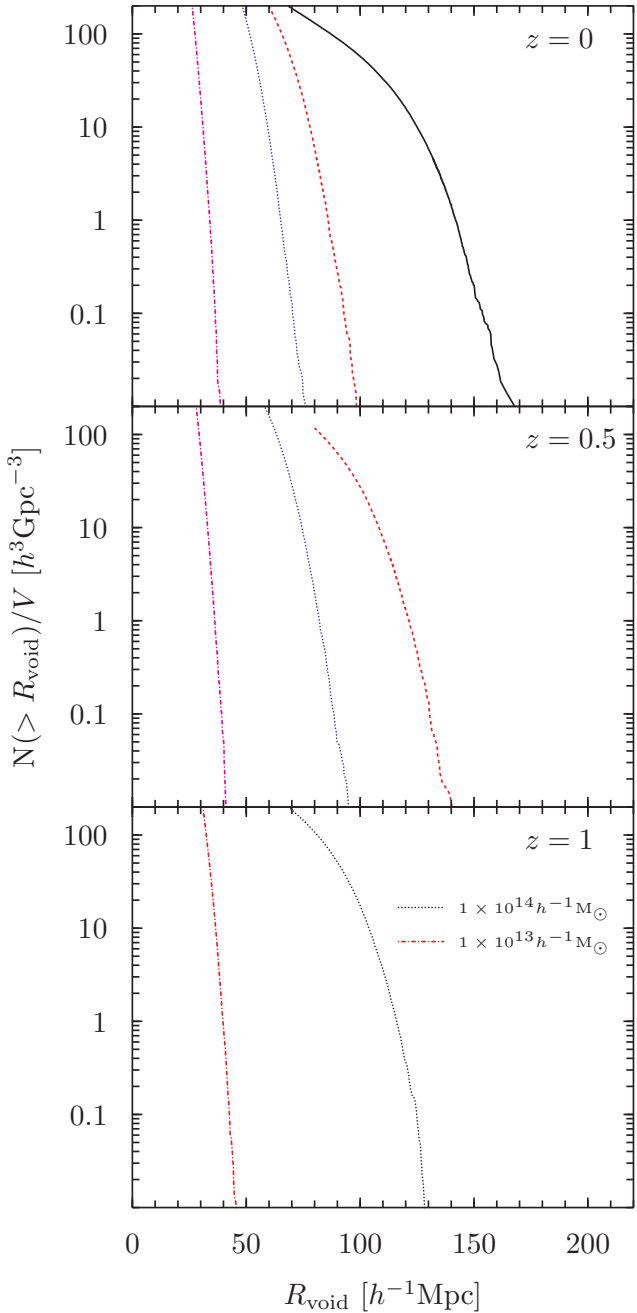}
\end{center}
\caption{Void functions from the Jubilee simulation for $z = 0, 0.5, 1$ top-to-bottom respectively. Voids are defined as spherical regions of radius $R_{\mathrm{void}}$ wherein no haloes with a mass higher than a threshold mass are found. The plot shows, for different threshold masses, the number densities of voids with radii over $R_{\mathrm{void}}$.}
\label{void:fig}
\end{figure}

In Figure~\ref{void:fig}, top panel, we show the void functions at $z = 0$ for four different threshold masses. For the largest threshold we find a few very large spheres with radii of 150$~h^{-1}$Mpc which do not contain any cluster more massive than $5\times10^{14}~h^{-1}\mathrm{M}_{\odot}$. For smaller thresholds the void function is very steep, i.e. there are a number of voids with a volume almost as large as the volume of the largest voids defined by the threshold. This means that the voids are almost uniformly distributed, as there are so many of a similar size. At higher redshifts (middle and bottom panel of Figure~\ref{void:fig}) we observe similar behaviour but, due to the evolution of the mass function, only with lower threshold masses. Note, that at the lowest threshold ($10^{13}~h^{-1}\mathrm{M}_{\odot}$) the maximum void radius is almost redshift independent between $z=0-1$ and occurs at a void radius of about 40$~h^{-1}$Mpc. This may seem in contradiction to the fact low density regions expand slightly faster than the mean expansion rate of the universe. However, since the tracers of the voids are also evolving the number of objects above the threshold evolves. For $10^{13}~h^{-1}\mathrm{M}_{\odot}$ mass haloes, the number counts rise from 38,994,056 at $z = 1$ to 91,615,821 at $z = 0$. Therefore, the mean distance shrinks from 18$~h^{-1}$Mpc to 13$~h^{-1}$Mpc and using this threshold mass we see the interesting result that the maximum void radius remains almost constant in time.

\section{Summary and Discussion}
\label{sect:disc}

In this paper we have presented a broad range of results from a $\Lambda$CDM-based simulation. The results have focussed on predictions on very large scales, such as extremely massive clusters and large void regions. The simulation itself represents one of the largest undertaken to-date, with a volume of $(6~h^{-1}\mathrm{Gpc})^3$, and haloes resolved down to $1.4\times10^{12}~h^{-1}\mathrm{M}_\odot$, a resolution that allows the creation of mock LRG and cluster catalogues.

The distribution of dark matter haloes in the Jubilee was found to be well described by fitting functions from the literature, and the dark matter haloes from the Jubilee have been used in a separate paper \citep{2013MNRAS.433.1230W} to construct mass function fits across a broad range of redshifts and volumes. For the rare tail of the mass function we have confirmed that the Poisson distribution describes well the number counts of objects. The masses of  clusters with extremal masses in the Jubilee simulation were investigated across a range of redshifts and were found to agree well with both observation and theory, in particular the expected masses of the very largest objects found when using Extreme Value Statistics.

\subsection{Implications for precision cosmology}

An important prediction from this simulation is the expected effect of cosmic variance on the counts of massive clusters. This result can be used to gauge  number-count errors in survey and simulation data. Understanding this is a vital component of the drive towards high precision cosmology. We showed in Figure~\ref{cosmic_variance:fig} how the expected number of clusters in given volumes are likely to vary. In general quantifying the effect of cosmic variance in simulations is notoriously tricky due to the requirement for either multiple repeats of a simulation or for a large simulated volume (or preferably both of these). Due to the large scale of the Jubilee volume we are able to use the latter and do so in a manner that includes the long-wavelength modes of the matter distribution. The variation in cluster counts for smaller boxes or surveys is highly significant if one is investigating the distribution of high-mass objects such as galaxy clusters, which form an important cosmological probe.

\subsection{The largest voids}

Our largest void, defined using a threshold mass of $5\times10^{14}~h^{-1}\mathrm{M}_\odot$, is $\sim350~h^{-1}\mathrm{Mpc}$ across. To put this void in context it is around one fifth of the volume of the Millennium simulation and it contains no clusters with mass greater than $5\times10^{14}~h^{-1}\mathrm{M}_\odot$. The probability, based on volume-occupation alone, of finding yourself within this void in the universe represented by the Jubilee simulation is $0.01\%$. There have been investigations as to whether our occupying a local underdensity might explain the apparent existence of an accelerated expansion in the late-time universe \citep{1979GReGr..11..281E,1997MNRAS.292..817M,1998ApJ...503..483Z,2001MNRAS.326..287T,2002PThPh.108..809I,2005PhRvD..71f3537B,2005gr.qc.....3099W,2005JCAP...10..012M,2009JCAP...09..025A,2010MNRAS.405.2231F,2010JCAP...12..021M,2011PhRvD..83f3506N}. The void in question would need to have very specific characteristics that include its radius, sphericity, density and density profile. Predictions for these void parameters vary but have typically required the void to be of at least a few hundred Mpc in radius and, importantly, close to spherically symmetric, with us as observers very near its centre. This latter requirement is due to the type Ia supernovae data implying that dark energy is close to isotropic across the sky. We see from our void functions in Figure~\ref{void:fig} that there are a few hundred voids in the Jubilee volume with radii $R_{void} > 100~h^{-1}\mathrm{Mpc}$, for the $5\times10^{14}~h^{-1}\mathrm{M}_\odot$ mass threshold. We estimate the proportion of the entire simulation volume taken up by voids with a radius of $R_{void} > 100~h^{-1}\mathrm{Mpc}$ to be 0.04\%. Adding an additional requirement that an observer occupy the central $1\%$ of the void volumes in question we arrive at the total spatial volume in the Jubilee box that would contain observers in the centre of voids of radii greater than $100~h^{-1}\mathrm{Mpc}$ to be $~\sim0.0004\%$. This is a rough statistical estimate and ignores the fact that observers might be better considered to only exist at the locations of galaxies in the simulation. In addition the simulation contains a dark energy component so has already modelled the effect of late-time accelerated expansion on structure formation. This latter point does not alter the order of magnitude of the result as void sizes in universes without dark energy are comparable to void sizes in $\Lambda$CDM \citep{2000MNRAS.318..280M}. We intend to look more closely into putting a probability on this figure for voids in the Jubilee simulation in a follow-up paper. 

\subsection{The $\Lambda$CDM model versus observations}

The distribution of most-massive clusters in the Jubilee was found to be in line with current theoretical predictions based on Extreme Value Statistics. The nature of Extreme Value Statistics is that it lacks predictive power in terms of constraining models, but it is a powerful method for ruling out models based on only a handful of extreme data points \citep[various authors have previously implemented studies in cosmology based on it, for example][]{1988MNRAS.231..125C, 2009EL.....8859001A, 2011MNRAS.414.2436C, 2011MNRAS.413.2087D,Harrison2011}. Had the masses of observed clusters in \cite{2012arXiv1210.4369H} lain significantly away from the expected EVS prediction then the $\Lambda$CDM model would be immediately placed in doubt. One result in this paper, that of the extreme nature of the Bullet cluster, is suggestive of a possible tension with $\Lambda$CDM. An Extreme Value Statistics approach is likely to cast this result in a more comprehensive context, but it is beyond the scope of this paper to attempt analysis along these lines.

In follow-up work we intend to investigate in more detail the existing discrepancy between observations of the ISW signal and the expected $\Lambda$CDM signal. For example \cite{2010MNRAS.401..547H} claim that the observed voids from SDSS data are too large for a $\Lambda$CDM universe. This result was based largely on analysis of the ISW signal in \cite{2008ApJ...683L..99G} and \cite{2009ApJ...701..414G}. This highlights the intimate link between the void distribution and the ISW signal -- underdense regions imprint themselves on the CMB via the ISW effect -- and represents a current challenge to the $\Lambda$CDM model.

\section*{Acknowledgments} The simulation was performed on the Juropa supercomputer of the J\"ulich Supercomputing Centre (JSC).  We would like to give special thanks to Alexander Schnurpfeil as well the whole the J\"ulich support team for their help during the phase of code implementation as well as during the production runs. We thank Peter Coles, Aurel Schneider and Leonidas Christodoulou for their helpful comments and feedback on this work. We also thank Steffen Knollman for his help in applying the AHF halofinder to our data. We thank Volker Springel for allowing us to use the Gadget-3 code with its FOF halofinder. WW thanks The Southeast Physics Network (SEPNet) for providing funding for his research. ITI was supported by The SEPNet and the Science and Technology Facilities Council grants ST/F002858/1 and ST/I000976/1. AK is supported by the {\it Spanish Ministerio de Ciencia e Innovaci\'on} (MICINN) in Spain through the Ram\'{o}n y Cajal programme as well as the grants AYA 2009-13875-C03-02, AYA2009-12792-C03-03, CSD2009-00064, CAM S2009/ESP-1496 and the {\it Ministerio de Econom'a y Competitividad} (MINECO) through grant AYA2012-31101. GY acknowledges support from MINECO (Spain) under research grants AYA2009-13875-C03-02, FPA2009-08958, AYA2012-31101 and Consolider Ingenio SyeC CSD2007-0050. He also thanks Comunidad de Madrid for support under PRICIT project ASTROMADRID (S2009/ESP-146). JMD and EMG acknowledge support of the consolider project CSD2010-00064 funded by the Ministerio de Economia y Competitividad and AYA2012-39475-C02-01.

\bibliographystyle{mn}



\end{document}